\documentclass[traditabstract]{aa} 
\usepackage{graphicx}
 \usepackage{lscape}
 \usepackage{colortbl}
 \usepackage{natbib}
 \usepackage{verbatim} 
  \usepackage{comment} 
  \usepackage{aalongtable}
  \usepackage[margin=2cm]{geometry}
\usepackage[font=small,labelfont=bf,tableposition=top]{caption}
\usepackage{hyperref}
 \bibpunct{(}{)}{;}{a}{}{,}
\usepackage{txfonts}
\begin{document}
   \title{Revisiting the Hubble sequence in the SDSS DR7 spectroscopic sample: a publicly available Bayesian automated classification}
\titlerunning{DR7 morphological classification}

   \author{M. Huertas-Company
          \inst{1,2}
          \and 
          J.A.L Aguerri
          \inst{3}
          \and
          M. Bernardi
          \inst{4}
          \and
          S. Mei
          \inst{1,2} 
          \and
          J. S\'anchez Almeida
          \inst{3}
          }

   \institute{GEPI, Paris-Meudon Observatory 5, Place Jules Janssen, 92190, Meudon, France\\
              \email{marc.huertas@obspm.fr}
         \and
             Universit\'e Paris Diderot, 75205 Paris Cedex 13, France
        \and
         Instituto de Astrof\'isica de Canarias, C/ V\'ia L\'actea s/n, 38200 La Laguna, Spain
         \and
        Department of Physics \& Astronomy, University of Pennsylvania, 209 S. 33rd St., Philadelphia, PA 19104, USA\\}

   \date{Received September 15, 1996; accepted March 16, 1997}

 
  \abstract
   {
    We present an automated morphological classification in 4 types (E,S0,Sab,Scd) of $\sim700.000$ galaxies from the SDSS DR7 spectroscopic sample based on support vector machines. The main new property of the classification is that we associate  a probability to each galaxy of being in the four morphological classes instead of assigning a single class. The classification is therefore better adapted to nature where we expect a continuous transition between different morphological types. The algorithm is trained with a visual classification and then compared to several independent visual classifications including the Galaxy Zoo first-release catalog. We find a very good correlation between the automated classification and classical visual ones. The compiled catalog is intended for use in different applications and is therefore freely available thorugh a dedicated webpage \thanks{\url{http://gepicom04.obspm.fr/sdss\_morphology/Morphology\_2010.html}} and soon from the CasJobs database. }
   {}

   \keywords{Catalogs, Astronomical databases, Galaxies:evolution, Galaxies:formation, Galaxies:fundamental parameters}

   \maketitle
%

\section{Introduction}

Classification of objects is a key step in understanding and analyzing an astrophysical sample. In particular, morphology is a powerful tracer of the structure of a galaxy. Since Hubble's first classification of galaxies according to their shape~\citep{Hubble26}, it has been shown that this phenomenological description hides important physical differences between galaxies and probably different evolutionary tracks. Elliptical galaxies appear with old stellar populations, high velocity dispersion, and small fraction of gas while spiral galaxies are more gas-rich, with younger stellar populations whose motion is rotation dominated.

The main problem with morphology comes from estimation, since, even when done through visual inspection, there are several intrinsic problems that can hardly be overcome. First, when one goes at high redshift, several new galaxies appear that do not necessarily fit in the Hubble fork (e.g. \citealp{abraham94,abraham96,conselice08,Delgado10}), and secondly, everybody who has looked at galaxies in detail has realized how difficult it is to classify them by eye since there are lots of objects that do not fall in a clear \emph{box} (e.g. \citealp{postman05}). This becomes even worse when other parameters are included such as colors or stellar dynamics. For example, \cite{Schawinski09} and \cite{Kannappan09} have found a significant fraction of elliptical galaxies with blue colors in the local Universe. In the SAURON project (e.g.~\citealp{Emsellem07}), one of the main conclusions is that a significant fraction of morphologically defined early-type galaxies present features similar to late-type ones, such as rotation in their cores. The definition of an early or late type galaxy is consequently not very clear. What defines a given galaxy type? Is it just a shape and bulge fraction? or is it shape and stellar populations? or is it stellar dynamics? Almost eighty years after Hubble's definition, these questions remain unanswered. 

It seems that, instead of being a closed definition, there is more like a continuous population of galaxies with some \emph{canonical} objects, prototypes of elliptical, or spiral galaxies and then some galaxies that are more or less close to the definition. Consequently, it makes more sense to assign distances or probabilities of being in one of the canonical classes instead of having a binary definition that is not necessarily very close to reality. 

In addition to these intrinsic issues, there are methodological problems as well because morphological classifications are, by definition, done by visual inspection. This job can be done on small samples but becomes an impossible task in very large surveys such as the SDSS, unless it is done through the aggregated efforts of hundreds of thousands of people over the course of many months as for the Galaxy Zoo project \citep{Lintott08, lintott10}.

 Lots of effort has been made to try to determine morphology in an automated and simple way by measuring some parameters, such as concentration, asymmetry, clumpiness, Gini index (e.g. \citealp{abraham96,Con00,abraham03}) or through 1D \citep{prieto01,trujillo01} or 2D-fitting algorithms (e.g. \citealp{Simard02,Peng02,deSouza04,mendez-abreu08}).
More sophisticated classifications include colors and color gradients (e.g.~\citealp{Neichel08}) or use neural networks (e.g. \citealp{Ball04}); however, all these methods deal with a finite number of classes and/or at some point require a degree of human intervention. Moreover, one can still argue that automated classifications are not \emph{real} morphological classifications since we are just measuring parameters of the light distribution while morphology is a much more complex pattern recognition problem. 

In \cite{huertas-company08,huertas-company09} we presented a method based on support vector machines (galSVM). 
It was initially designed for high-redshift galaxies, and it has the advantage of dealing with an unlimited number of parameters and assigning probabilities instead of binary classes. We showed that, when applied to poorly resolved samples, it increases the accuracy by a factor of $\sim3$, compared to more classical methods. The method has already been used and validated in a variety of different cases on space and ground-based data  to study, for instance, the fraction of blue early-type galaxies in the field \citep{huertas-company10} and the morphological mixing in clusters at intermediate redshift \citep{huertas-company09b}. 

In this paper, we revisit the Hubble sequence in the SDSS DR7 spectroscopic sample using this method and assign a probability to each galaxy of being in the following morphological classes: E,S0,Sab,Scd, instead of a closed class. 
The paper proceeds as follow. In section~\ref{sec:sample} we describe the sample used, and in section~\ref{sec:method} the method employed for the classification is presented in detail. We discuss the robustness of the classification at the faint end in section~\ref{sec:disc} and a comparison with a detailed visual classification of $\sim$14000 galaxies \citep{Nair10} and with the Galaxy Zoo first release catalog \citep{lintott10} is shown in section~\ref{sec:comp}. Finally, we show some examples of how to use this catalog in section~\ref{sec:examples}.

\section{The sample}
\label{sec:sample}
We used all the SDSS DR7 spectroscopic sample as the starting base. Then, the selection of objects was based on \cite{sanchez-almeida10} who performed an unsupervised automated classification of all the SDSS spectra. Basically, we chose galaxies with redshift below 0.25, and with good photometric data and
clean spectra, meaning objects not too close to the edges, not saturated, or not properly deblended. The final catalog contains 698420 objects for which we estimate the morphology as shown below. No additional selection criteria were added so that the catalog is not biased to any particular application. 

\section{The method}
\label{sec:method}
The classification method is based on support vector machines (SVM) implemented in the libSVM library \citep{libSVM}. SVM is a machine learning algorithm that tries to find the optimal boundary (not necessarily linear) between several clouds of points in an N-dimensional space. More information about the algorithm can be found in \cite{huertas-company08}. There are several interesting properties that make this algorithm attractive for galaxy classification. First, it can deal with an unlimited number of dimensions so that everything that is related to the classes one would like to separate can be included in the classification process. Second, it does not deliver a binary classification but a probability of belonging to a given class. This probability is related to the accuracy of the classification, the higher it is, the higher the success rate (and so the closer are the objects to the \emph{canonical classes}), so that the accuracy of the classification can be studied in an objective way. This property is lacking in most of the existing classification schemes (specially in the visual techniques). 

\subsection{Training sample}
The SVM method needs a training sample, and all the behavior of the learning algorithm depends on how close this training sample is to the real sample one wants to classify. For morphological classification, the training sample is typically built using a visually classified subsample. The problem is that, usually, visual classifications are performed on the brightest objects because it is obviously easier, and one would like to go fainter in automated classifications. This causes a mismatch in the properties of the galaxies in the training sample and in the real sample, which can lead to misclassifications. One solution, as shown in \cite{huertas-company08}, is to simulate faint galaxies. In this paper, we decided not to include any simulations to be able to use the parameters measured in the SDSS database so as be consistent in the way parameters were measured in the training and real samples. The effects of such (risky) decisions are carefully studied in sections~\ref{sec:disc} and~\ref{sec:comp}. We therefore used \cite{fukugita07} classification as the training sample. In their paper, they provide a visual classification of 2253 SDSS galaxies brighter than $m_r=16$ (compared to the full DR7 sample, which goes up to $m_r\sim18$).  Since our goal is to classify galaxies in 4 main classes (E,S0,Sab,Scd), we group them according to their morphological index T (\citealp{fukugita07}, Table 1): E: $T<1$, S0: $T=1$, Sab: $2<T<4$, and Scd: $4\leq T<7$ before using them for training the algorithm. We included irregulars ($T =6$) in the Scd class since there are not enough objects in the local universe (and in particular in the \cite{fukugita07} catalog) to make a separate class for the training. 

\subsection{Procedure}
\begin{figure} 
\centering 
 \resizebox{\hsize}{!}{\includegraphics{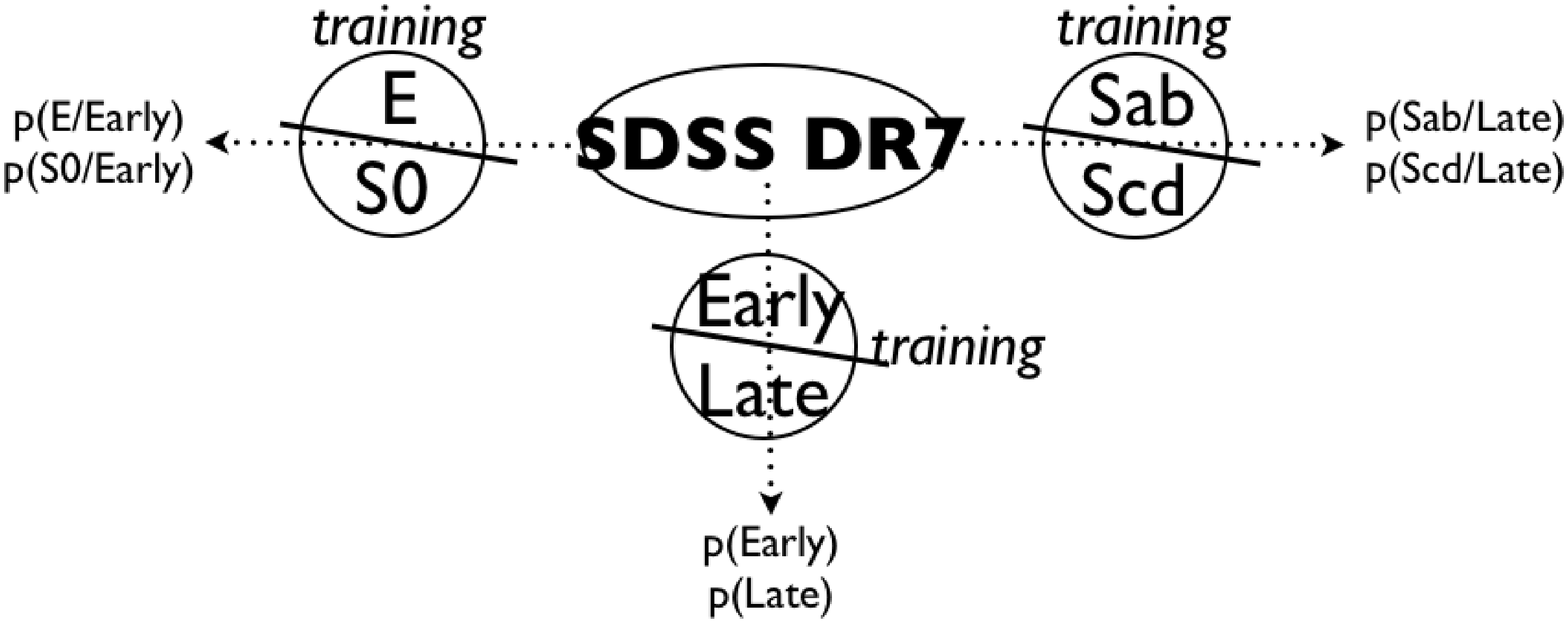}}
 \caption{Schematic view of the procedure used to classify the sample and the probabilities measured in each step.}.  
 \label{fig:proced} 
 \end{figure}

SVM were originally thought to separate 2 classes. Some implementations were done to add multi-class separation but the accuracy is more difficult to assess. To avoid dealing with multi class problems, in this paper we proceeded in two steps. First we separated the sample in two main classes, i.e. early-type galaxies, which includes ellipticals and S0 galaxies, and late-type galaxies, which contain all the remaining morphological types from Sa to Scd/Im. Then we took the whole sample and classified it again using 2 different training sets that contain only early-type and late-type galaxies respectively (see figure~\ref{fig:proced}). The probability computed in this second step can thus be seen as a conditional probability: ``probability of being S0 or E given that it is an early-type galaxy" and ``probability of being Sab or Scd given that it is a late-type galaxy". With this approach we were certain to have a broad classification in two types (which is enough for lots of science applications) with a high success rate, and then a more detailed one. Each galaxy in the catalog is therefore associated with 6 probability values, i.e. the probability of being in the two broad classes and the probability of being in the 4 subclasses. The 4 probabilities of the 4 subclasses can be computed with the Bayes theorem using the conditional probabilities:

\begin{equation}
P(E)=P(Early)\times P(E/Early) 
\end{equation}
\begin{equation}
P(S0)=P(Early)\times P(S0/Early) 
\end{equation}
\begin{equation}
P(Sab)=P(Late)\times P(Sab/Late) 
\end{equation}
\begin{equation}
P(Scd)=P(Late)\times P(Scd/Late) 
\end{equation}

We considered in the 4 equations above that  $P(Early/E)=P(Early/S0)=P(Late/Sab)=P(Late/Scd)=1$. Following these equations, we obviously have $P(Early)=P(E)+P(S0)$ and $P(Late)=P(Sab)+P(Scd)$ and P(E)+P(S0)+P(Sab)+P(Scd)=1. 

\subsection{Parameters used}
The SDSS database contains lots of photometric and spectroscopic parameters that are related to the morphological type of the galaxy and could hence be used for the classification. One interesting property of SVM is that they are not degenerate, in the sense that adding extra-parameters does not lead to a decrease in the classification accuracy \citep{huertas-company08} even if they do not bring any extra information. However, the computing time increases and the parameter space is less well sampled if too many parameters are included. After several tests, we decided to include three types of parameters: (1) color (g-r,r-i) k-corrected with \cite{Blanton05} code, (2) shape (\textsc{isoB/isoA} in the i--band and \textsc{deVAB\_i}), and (3) light concentration (\textsc{R90/R50} in the i--band). For color measurements we use model magnitudes corrected for {galactic extinction. \textsc{isoB} and \textsc{isoA} are the isophotal minor and major axes respectively, and \textsc{deVAB\_i} is the DeVaucouleurs fit \emph{b/a}.  \textsc{R90} and \textsc{R50} are the radii containing 90\% and 50\% of the petrosian flux, respectively. Adding more parameters does not significantly change the classification and increases the execution time. The decision to include the color could be discussed, since, as pointed out in the introduction, it is not clear how an early-type or a late-type galaxy is actually defined. Since our approach is to define classes as closely as possible to the \emph{canonical} definition and then compute \emph{distances} to them, it makes sense to include color. Indeed, for an elliptical to be elliptical it should be red, otherwise it should be called blue elliptical, and it is an exception to the normal classification. Eitherway, tests performed reveal that removing the color from the parameter space does not significantly change the classification. Fewer than $10\%$ of the galaxies change their main morphological class. In figure~\ref{fig:param_proba}, we show the 4 probabilities as a function of some representative parameters used in the classification. We observe some obvious correlations: i.e. the probability of being elliptical increases with concentration, and redder galaxies have higher probabilities of being ellipticals. The correlations are less clear for intermediate classes (S0 and Sab). One important conclusion by looking at these plots is that one single parameter is not enough to select galaxies with high probability of being in a given class. For instance, it is common to use a concentration threshold
$R90/R50 > 2.6$ (in the r-band) to select elliptical galaxies
(e.g. \citealp{Bell03, Kauffmann03}). As shown in the top panel of figure~\ref{fig:param_proba} this selection results in a significant fraction of galaxies with low probabilities of being elliptical galaxies (as also shown in  \citealp{Bernardi10}). 

\begin{figure*} 
\centering 
 \resizebox{\hsize}{!}{\includegraphics{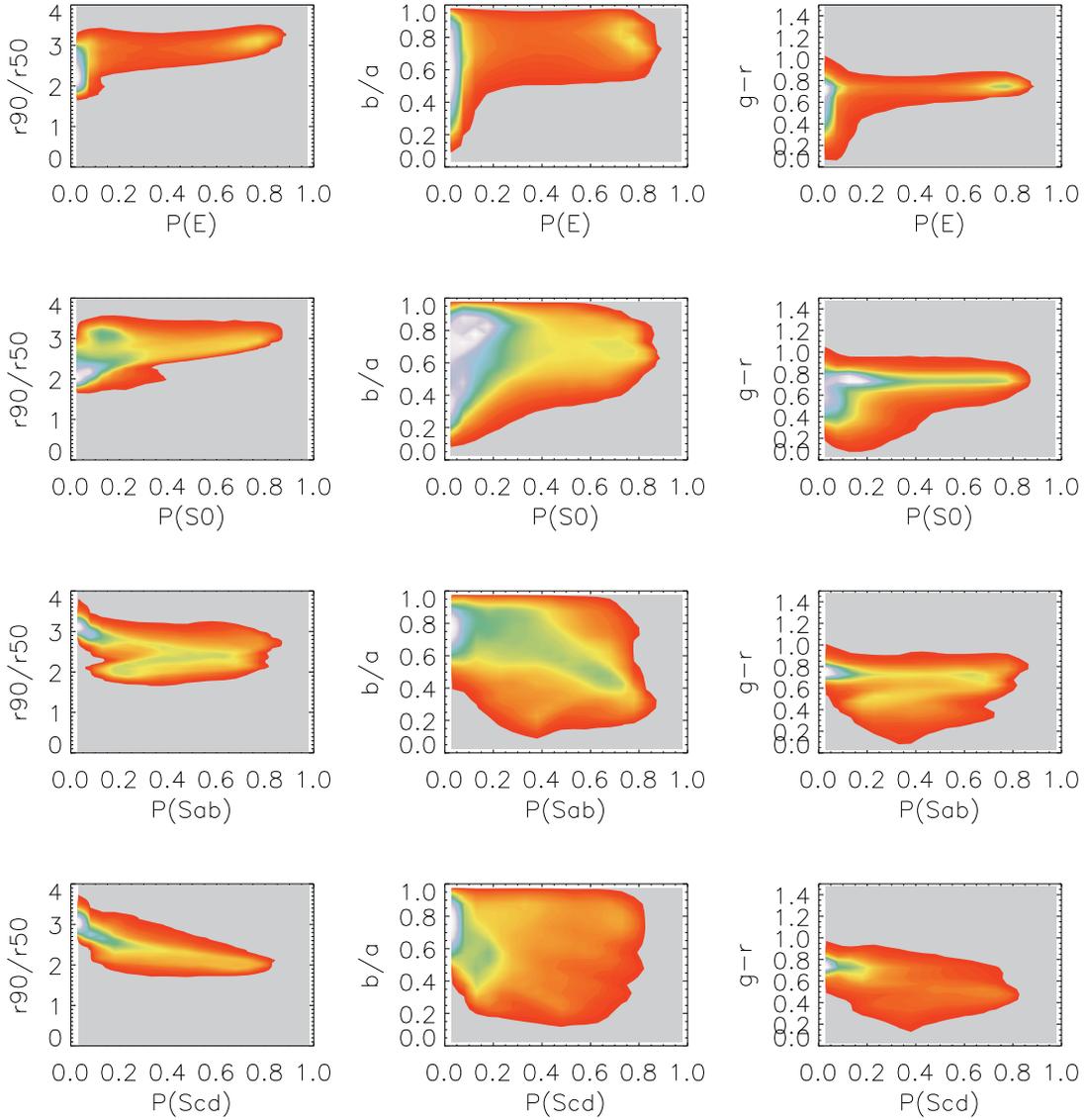}}
 \caption{ Distribution of the main parameters used (concentration (r90/r50), axis ratio (b/a), and color (g-r)) as a function of probability. All parameters are measured in the i band (see text).}.  
 \label{fig:param_proba} 
 \end{figure*}

\section{Robustness}
\label{sec:disc}
\subsection{Accuracy at the faint end}
As pointed out in section~\ref{sec:method}, there is a critical point in our approach, since the classified sample contains lots of galaxies fainter than the limiting magnitude of the training sample. Therefore, it is very important to check that these faint galaxies are not systematically misclassified just because they are not represented in the training. 
As a first check, we computed the probability distributions of bright ($m_g<16$) and faint galaxies ($m_g>16$) in figure~\ref{fig:proba_faint_bright} to check that faint galaxies are systematically classified with lower probabilities. As shown in \cite{huertas-company08}, the probability is a kind of measure of how good the classification is and how close a given galaxy is to the corresponding associated class. Low probabilities in all the classes consequently mean that the galaxy is not close to any of the classes of the training, which would mean that faint galaxies are not properly classified because they are not properly sampled in the training set. We observe in figure~\ref{fig:proba_faint_bright} that there is no evident difference between both probability distributions. A Kolmogorov-Smirnoff test gives between 99\% and 55\% probability that the 2 distributions are drawn from the same distribution, so the possibility that the 2 distributions are decoupled is rejected.
The probability values seem to be quite independent of the galaxy brightness, at least up to the magnitude limit of the sample. The algorithm is thus able to find a clear, closest class even for the faintest objects, which supports the robustness of the classification. \\

As a second check, we looked at some of the images of the faint end of the sample (Fig.~\ref{fig:gal_examples}). We confirm that high-probability values for a given morphological class still correspond to galaxies that closely look like galaxies in this given class independently of the magnitude. It therefore seems that the classification is robust even for the faintest objects in the sample and that no major misclassifications are evident. In section~\ref{sec:comp} we perform a detailed comparison with a visual classification of faint objects. 

 \begin{figure}[h!] 
\centering 
 \resizebox{\hsize}{!}{\includegraphics{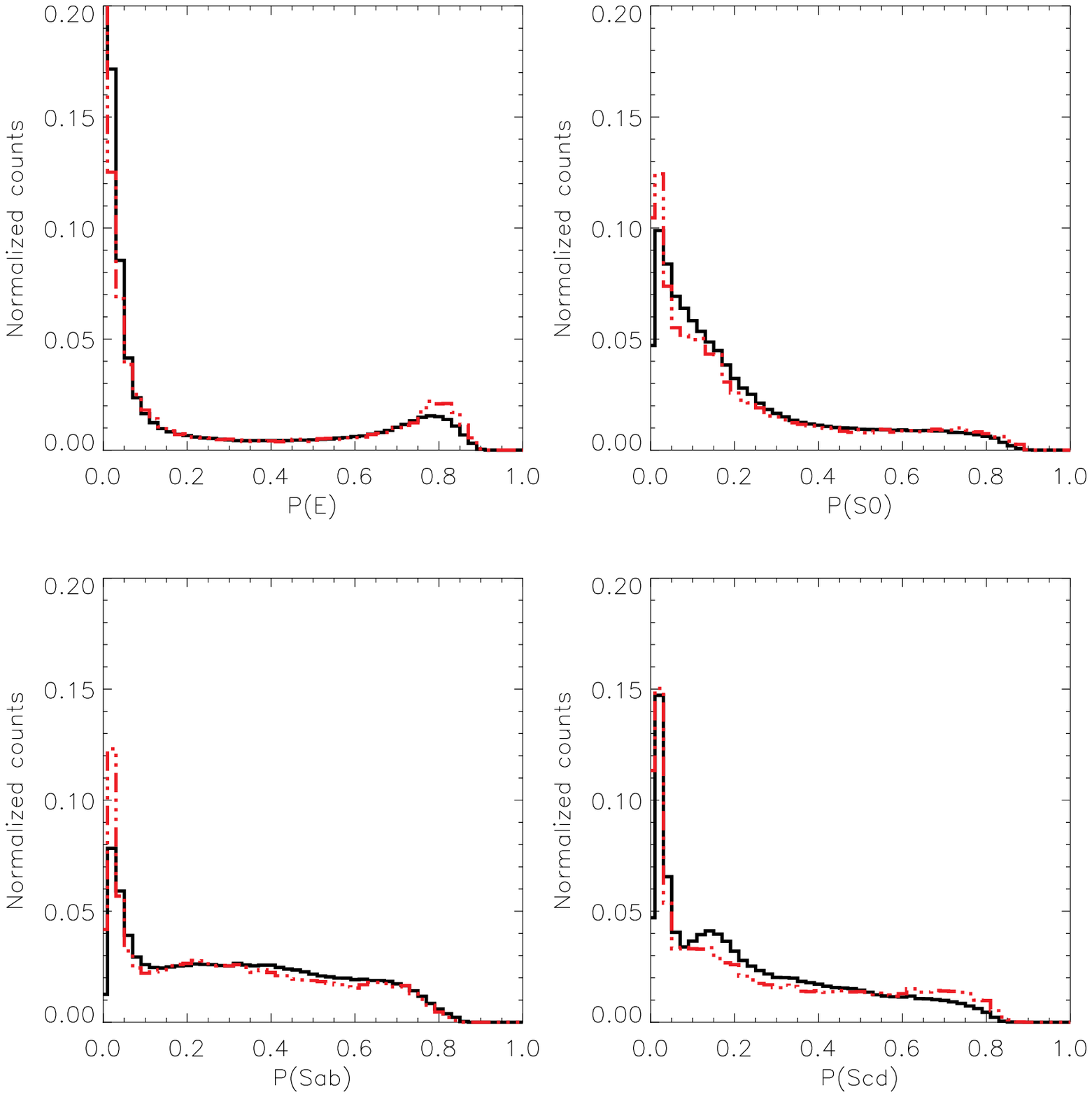}}
 \caption{Probability distributions of bright ($m_g<16$, red dotted line) and faint ($m_g>16$, black solid line) galaxies in the sample. The 4 panels show the 4 computed probabilities as indicated in the x-axis labels.}.  
 \label{fig:proba_faint_bright} 
 \end{figure}

\begin{figure*} 
\centering 
\resizebox{\hsize}{!}{\includegraphics{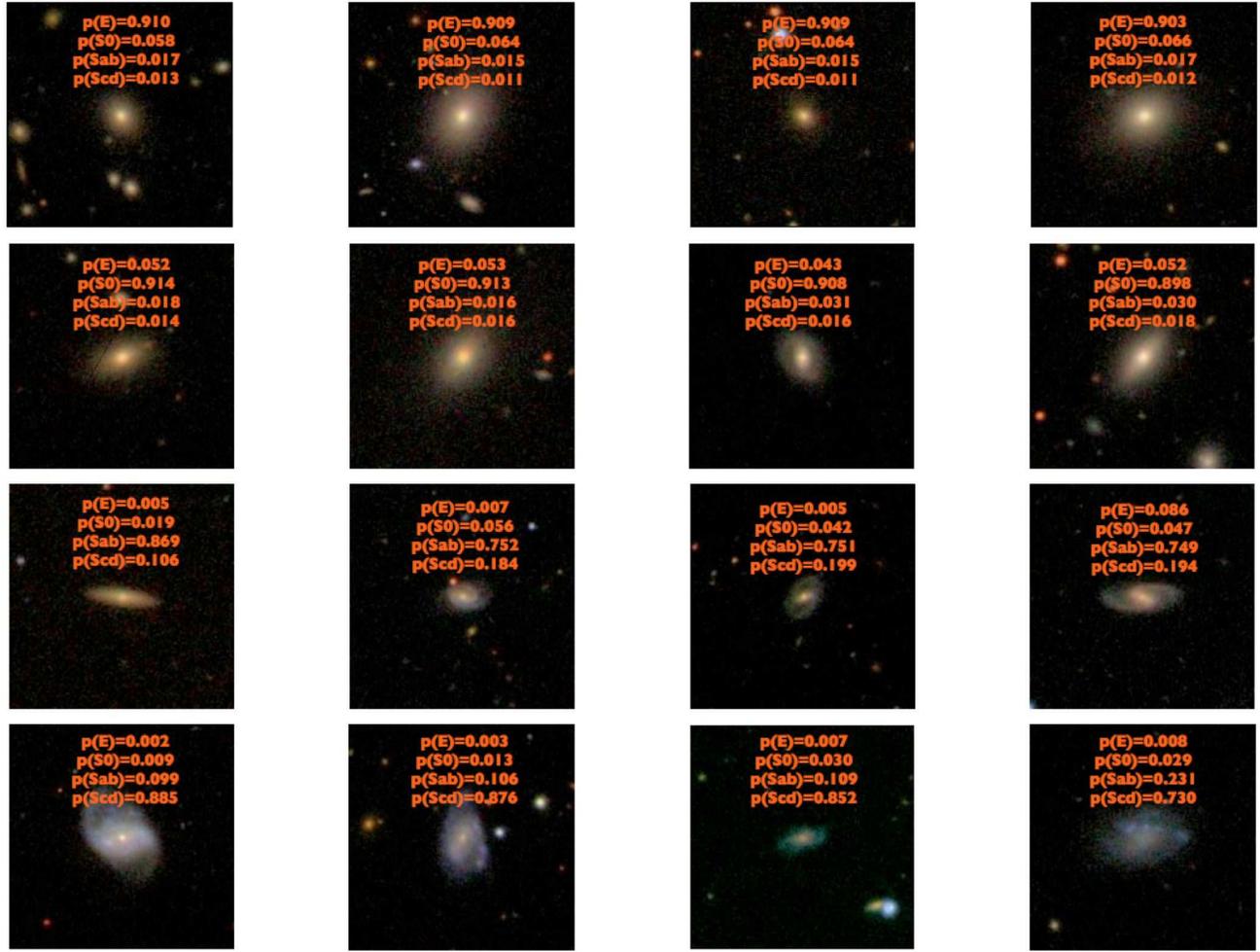}}
 \caption{Examples of galaxies with their computed probability values}.  
 \label{fig:gal_examples} 
 \end{figure*}

\subsection{Dependence on the training set}
Another important point that should be studied is the effect of changes in the training set on the final classification. In fact, a robust classification should not change significantly if some elements are removed from the training sample. On the contrary, if removing some elements leads to a completely different classification, it means that the parameter space is not properly sampled and therefore the classification is very unstable. To check this point, we performed 10 different classifications with slightly different training sets. The samples were generated by randomly selecting a subset of 500 galaxies from the \cite{fukugita07} sample. We then compared the different classifications in terms of probability. These 10 runs on the full data set take only a few minutes on a normal laptop.The average scatter over the 10 runs of the probability of being early-type (or late-type) is $12\%$.  In other words, when one changes the training set, the probability for a given galaxy changes $\sim12\%$ on average. This 12\% scatter is compatible and even less than the typical scatter found when several people perform visual classifications on the same sample (e.g.\citealp{postman05,fukugita07}).

\subsection{Uncertain objects}

Another way of assessing the robustness of the classification is by measuring the fraction of objects whose classification is uncertain. If this fraction appears to be too high it would imply that the algorithm is not working for a large fraction of the sample. We define \emph{uncertain} objects as those for which the difference between the maximum and the minimum probability value is less than 0.15; i.e. the four probabilities are in a range less than 0.15, so the galaxy does not clearly fit in any of the four morphological classes. 


There are 3013 objects verifying this condition, 0.4\% of the whole sample. The vast majority of the objects are therefore close to one (or two) morphological classes and very few are in an \emph{uncertain region}. A visual inspection of these galaxies (fig.~\ref{fig:uncertain}) reveals that they are small, compact, and/or disturbed objects, for which the visual morphology is also difficult to assess. They are not, however, particularly distant or faint objects since the magnitude and redshift distributions are compatible with the ones of the full sample. 

 \begin{figure}[h!] 
\centering 
 \resizebox{\hsize}{!}{\includegraphics{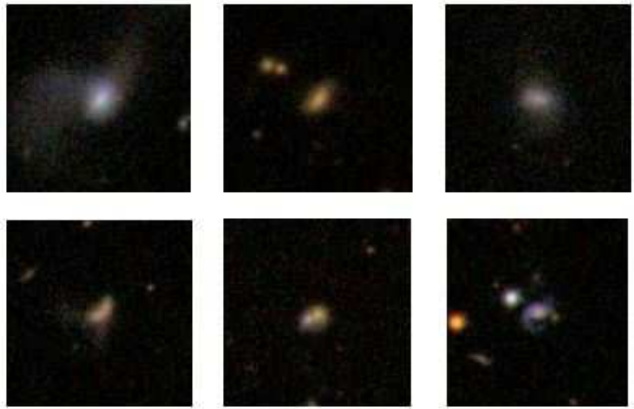}}
 \caption{Examples of uncertain classifications as defined in the text.}.  
 \label{fig:uncertain} 
 \end{figure}

\section{Comparison with visual classifications}
\label{sec:comp}
\subsection{Comparison with Nair \& Abraham 2010}
One obvious validation check of the classification is to compare it with existing visual classifications. As explained in previous sections, we used the \cite{fukugita07} catalog for training. It is therefore better to use a different independent subsample for testing the accuracy and robustness of the classification. In a recent paper, \cite{Nair10} published a very detailed visual catalog of 14034 galaxies in the SDSS with $m_g<16$. Galaxies in this sample are included in our classification, but most of them have not been used to build our training sample so they represent an ideal independent cross check. 
Since \cite{Nair10} classification is much more detailed than ours, we group their classes into 4 groups matching the 4 classes we have defined in this work. We consider elliptical galaxies objects with $TType=-5$, S0s, $TType=-2$, Sabs, $1\leq TType\leq 3$, and finally Scd, $5\leq TType\leq10$ (see table 1 of \citealp{Nair10} for a definition of the TType index used in their work). Figure~\ref{fig:Nair_SVM} shows the probability distributions of these 4 groups. Globally, we observe a good correlation between the probability values and the visual class. For example, galaxies visually classified as ellipticals have on average a probability of $\sim0.8$ of being ellipticals and $\sim0.2$ of being S0. The two other probabilities are almost zero. Traditionally, it is well known that it is very difficult to separate S0 galaxies by eye. This is reflected in the probability distributions which are more uniform than for the pure elliptical class. A galaxy visually classified as S0 has on average $\sim0.4$ probability of being S0 but also $\sim0.32$ of being elliptical and $0.2$ of being Sab, which reflects the difficulty of defining the S0 class and the fact that these galaxies are indeed a transition class in terms of morphology between the ellipticals and the spirals. A similar effect is seen in the Sab population which has on average a probability of $\sim0.55$ of being Sab but also $\sim0.15$ of being S0 or Scd. 
Another interesting measurement is the fraction of \emph{catastrophic} classifications, i.e. galaxies whose automated and visual classes are completely different. We define those cases as objects for which $P(E)>0.8$ and $TType > 5$ or   $P(Scd)>0.8$ and $TType = -5$, i.e. galaxies that are clearly elliptical (Scd) for our algorithm and visually classified as Sc or later (elliptical). There are only 2 objects verifying these conditions, and both are in the first case. They are indeed spiral galaxies, so the algorithm is wrong, but both have a large red bulge, which can probably account for the misclassification.

\begin{figure*} 
\centering 
 \resizebox{\hsize}{!}{\includegraphics{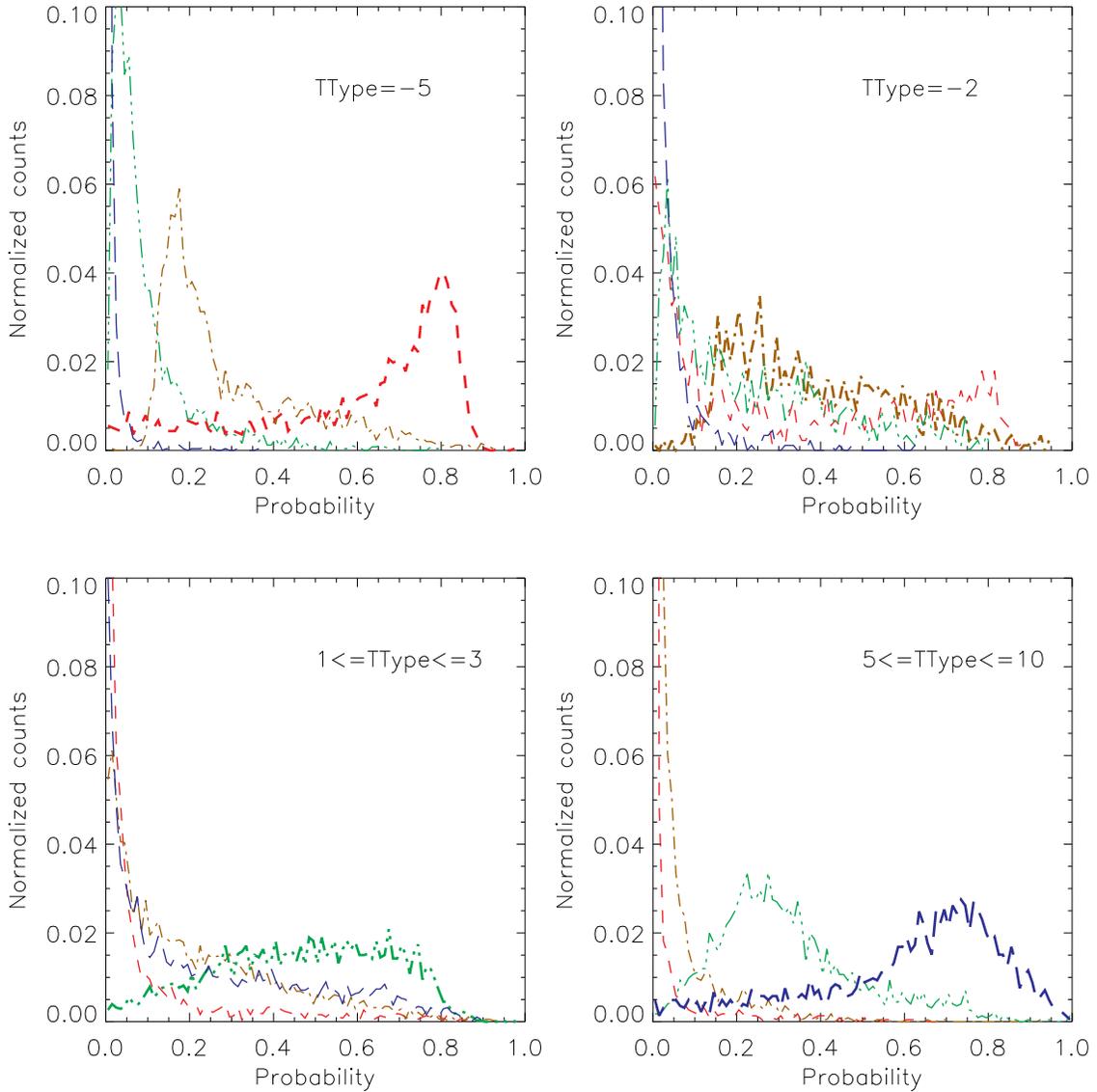}}
 \caption{Probability distributions of the 4 morphological types considered in this work for 4 visual types (TType) from \cite{Nair10}. Each different panel shows a different visual type. Top left panel shows galaxies with $TType=-5$ (Ellpticals); top right panel: $TType=-2$ (S0s); bottom left panel: $1\leq TType\leq 3$ (Sabs) and finally bottom right panel: $5\leq TType\leq10$ (Scd). Red short dashed lines are P(E), orange dashed dotted lines are P(S0), green dashed three dotted lines are P(Sab), and blue long dashed lines are P(Scd).  See text for details of how these 4 probabilities are computed. }.  
 \label{fig:Nair_SVM} 
 \end{figure*}

\subsection{Galaxy Zoo}
Recently, the Galaxy Zoo\footnote{http://galaxyzoo.org/} team \citep{lintott10} has made publicly available the visual classification of the full DR7 performed through the aggregated efforts of hundreds of thousands of people over the course of many months. This work is an extraordinary effort (and probably the only way) to visually classify present and future extremely large surveys. The main drawback, however, is that it requires plenty of time (more than 2 years in this case) to collect all the information and put all the catalogs in place. It is therefore a very interesting question to see how our automated classification behaves compared to this visual classification. Our classification is indeed much faster and can be run several times with different parameters in just a few minutes, but it is not obvious whether we can reach an accuracy similar to the human brain. Moreover, this comparison also enables the comparison for the faint end of the sample (since the GalaxyZoo catalog contains all galaxies), hence a new evaluation of the effect of lacking faint objects in the training sample (see section~\ref{sec:disc}). The classification made in the framework of the GalaxyZoo is less detailed than a \emph{pure} visual classification, such as the one from \cite{Nair10} or \cite{fukugita07}; i.e, they basically asked people if the galaxy is elliptical like (which should include S0s) or spiral like (with different subcategories like clockwise or anti-clockwise rotation), but without submorphological types. Galaxy Zoo 2\footnote{http://zoo2.galaxyzoo.org/} and Hubble Zoo\footnote{http://hubble.galaxyzoo.org/} will furnish more detailed classifications in the coming future but are not publicly available for the moment. The confidence of the classification in the current release is measured by the fraction of votes received, since each galaxy is classified by several persons. A galaxy is then flagged as early-type or spiral-like if the fraction of votes in one of those categories is greater than 80\%. In figure~\ref{fig:comp_galzoo1} we show the probability distribution obtained with the galSVM classification for galaxies flagged as elliptical like (flag ELLIPTICAL = 1) and spiral like (flag SPIRAL = 1), respectively. We observe an extremely good correlation between both classifications even for faint galaxies not necessarily well represented in the training set as discussed in \S~\ref{sec:disc}. Galaxies flagged as ellipticals in the Galaxy Zoo catalog have a median probability of 0.92 of being elliptical or S0 and the same for galaxies classified as spirals. This means that \emph{robust} classifications in Galaxy Zoo are also very sure classifications in our catalog; however the fraction of galaxies without a clear morphological type (i.e. the fraction of votes is less than $80\%$ so they lie somewhere between a pure early-type or late-type galaxy) in the Galaxy Zoo is relatively high ($\sim60\%$), so it is interesting to check where all these remaining galaxies fall.\\
For that purpose, we push the comparison a bit further. As a matter of fact, since the quality of the classification in the Galaxy Zoo is measured by the number of votes, another interesting test is to compare our probability measurement to the fraction of votes. In other words: does the probability measurement reflect the choice of the majority? We indeed expect to find a correlation, since \emph{certain} classifications in terms of votes should also be galaxies close to the \emph{canonical} definition, hence objects with high probability values. This comparison is shown in fig.~\ref{fig:comp_galzoo2}. There issignificant scatter, but we observe 2 clear clouds. Objects with a high fraction of votes for being elliptical have high probability values and vice-versa. The same behavior is measured for spirals. When we average the fraction of votes per probability bin, the correlation becomes clearer, and we find that there is a monotonic relation between the fraction of votes and the probability (Fig.~\ref{fig:comp_galzoo2}). This fact confirms that our probability measurement indeed measures the robustness of the classification for a given object. \\
In figures~\ref{fig:comp_galzoo3} and \ref{fig:comp_galzoo4} we compare the fraction of votes with the 4 more detailed probabilities computed in this work (P(E),P(S0),P(Sab), and P(Scd)). We again find a clear correlation between the number of votes given by people and the probability computed in an automated way by galSVM. 
 
\begin{figure*} 
 \resizebox{\hsize}{!}{\includegraphics{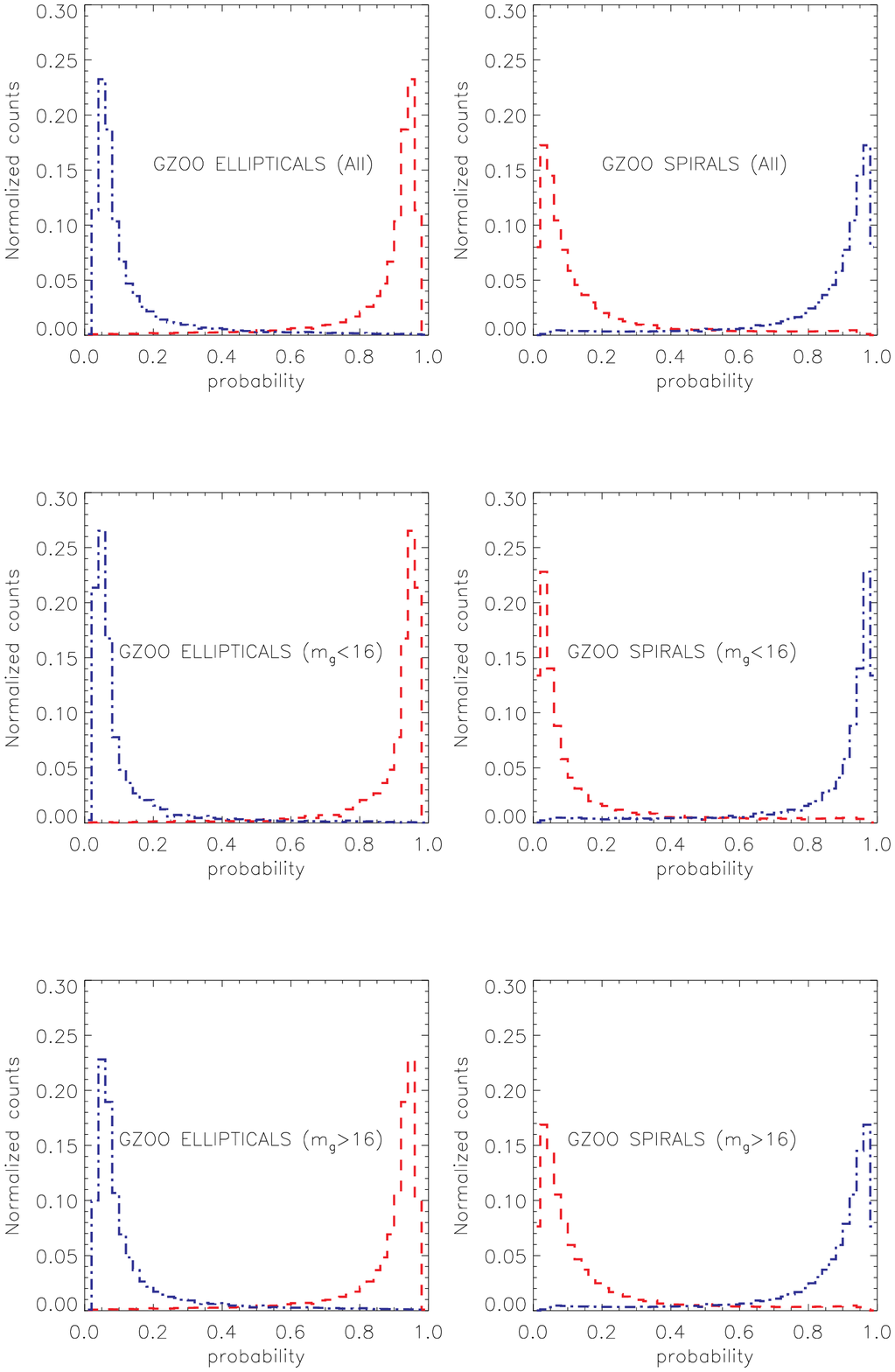}}
 \caption{Probability distributions of p(Early), red dashed line, and p(Late), blue dashed dotted line for galaxies flagged as ellipticals (left column) and spirals (right column) in the Galaxy Zoo. Top line shows all galaxies, and middle and bottom lines show bright ($m_g<16$) and faint ($m_g>16$) galaxies, respectively.}  
 \label{fig:comp_galzoo1} 
 \end{figure*}

\begin{figure*} 
 \resizebox{\hsize}{!}{\includegraphics{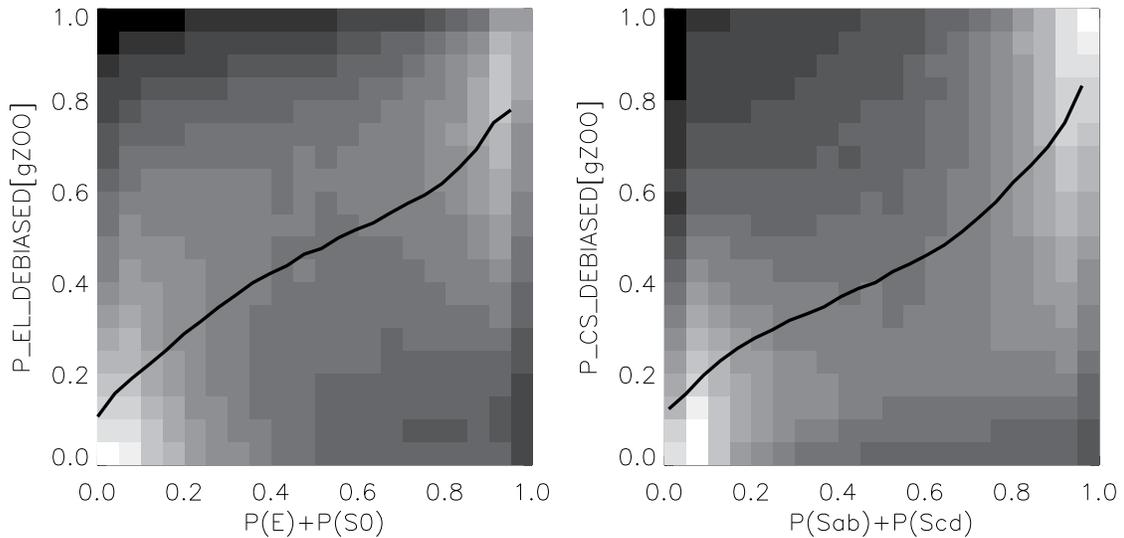}}
 \caption{Probability values computed with galSVM compared to the fraction of votes for ellipticals (left pannel) and spirals (right pannel). Gray scales are scaled to the data; i.e. white is maximum and black is minimum. Lines show the average fraction of votes in 0.05 probability bins}.  
 \label{fig:comp_galzoo2} 
 \end{figure*}

\begin{figure*} 
 \resizebox{\hsize}{!}{\includegraphics{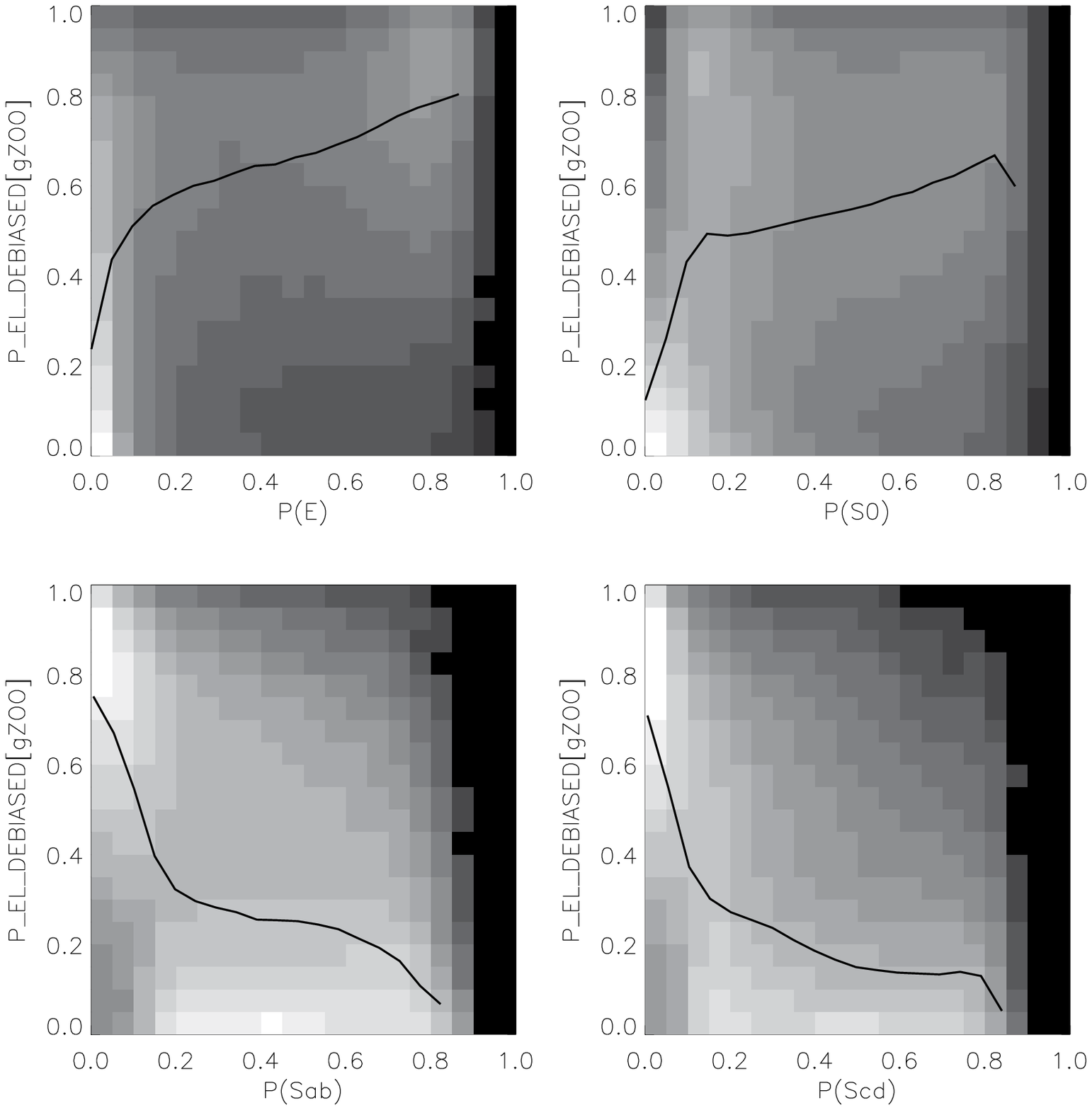}}
 \caption{Comparison between the fraction of votes for a galaxy to be ellipical like from Galaxy Zoo and the computed probabilities in this work. Gray scales are scaled to the data; i.e. white is maximum and black is minimum. Solid line shows the average relation. The average is computed in 0.05 probability bins.}.  
 \label{fig:comp_galzoo3} 
 \end{figure*}

\begin{figure*} 
 \resizebox{\hsize}{!}{\includegraphics{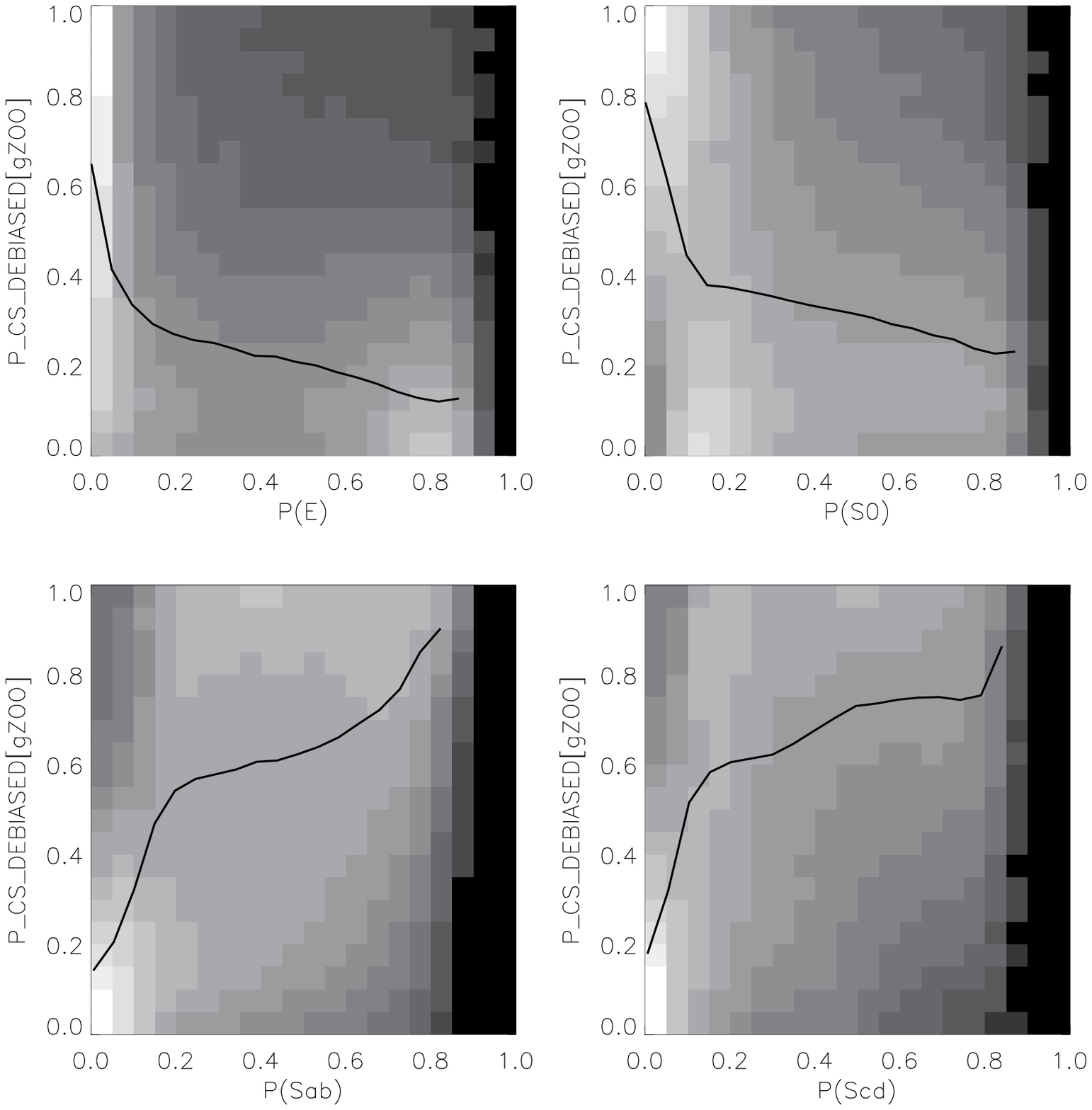}}
 \caption{Comparison between the fraction of votes for a galaxy to be spiral like from Galaxy Zoo and the computed probabilities in this work. Gray scales are scaled to the data, i.e. white is maximum and black is minimum. Solid line shows the average relation. The average is computed in 0.05 probability bins.}.  
 \label{fig:comp_galzoo4} 
 \end{figure*}

\section{How to use the catalog?}
\label{sec:examples}
The most important new point of the classification presented in this work is the measurement of probabilities. Therefore, a morphological class is not defined as a closed box, but there is more like a continuous transition from one class to another. How can this new property can be used for selecting a particular population and studying its properties? 
If one wants to perform luminosity or mass functions for a given morphological type, the optimal way (in terms of optimal estimation) is to make use of the probability measure as a weight for the galaxy counts. As shown in \cite{huertas-company09}, we can define a random variable $Y_k$: \[ Y_k = \left\{ \begin{array}{ll}
0 & \mbox{with a probability $1-P_{Type}$}\\
1 & \mbox{with a probability $P_{Type}$} \end{array} \right. \]. This way, the number of galaxies of a given morphological type in a mass or luminosity bin is simply given by its mathematical expectation, 
\begin{equation}
N_{Type}=\sum_{Nobj}{P_{Type}},
\end{equation}
and the $1-\sigma$ error is the square root of the variance:
\begin{equation}
\sigma_{Type}=\sum_{Nobj}{P_{Type}\times(1-P_{Type})}. 
\end{equation}

All the galaxies contribute to the mass function of a given morphological type weighted by its probability. As a result, a galaxy that is 95\% Sd and 0.5\% E will still contribute to the mass function of elliptical galaxies with a weight of 0.005. 

Another approach is to make probability cuts. This way, we decide that galaxies belong to a given class by applying a probability threshold.This approach (even if not optimal) should be closer to the classical approach from visual classifications in which galaxies only contribute in one given class. The threshold to apply depends on the application. For example, it is interesting to determine which threshold is the best to get similar distributions than with visual classifications. In figure~\ref{fig:mass_counts_nair}, we compare the two estimations of the observed distribution of stellar masses with the ones obtained from the visual classification of \cite{Nair10}. We use a threshold of $P_{Ttype}>0.45$ in each type and obtain similar distributions for all morphological types. Stellar masses are taken from the \cite{Nair10} catalog, also taken from \cite{Kauffmann03} estimates.

In figure~\ref{fig:mass_counts_proba} we show the observed distribution of stellar masses for the whole sample for different morphological types using the probability estimator. In this case, stellar masses are computed with the \cite{Bell03} formula, adapted from \cite{Bernardi10} to account for evolution:
\begin{equation}
log_{10}(M_{*}^{Bell}/M_\odot)=1.097(g-r)-0.406-0.4(M_r-4.67)-0.19z.
\end{equation}
We observe the expected trend; i.e, the mass function peaks at lower values for later morphological types. In the same figure, we compare the distribution of masses obtained from the Galaxy Zoo classification. We compare the one obtained with galaxies flagged as ellipticals (FLAG ELLIPTICAL = 1) with the one obtained using the two estimators described above, i.e. galaxies having $p(E)>0.5$ and probability weighting. The same is computed for spirals. There is almost a perfect match with the distributions computed using galSVM, which again confirms the accuracy of the automated classification presented in this paper. 

\begin{figure*} 
 \resizebox{\hsize}{!}{\includegraphics{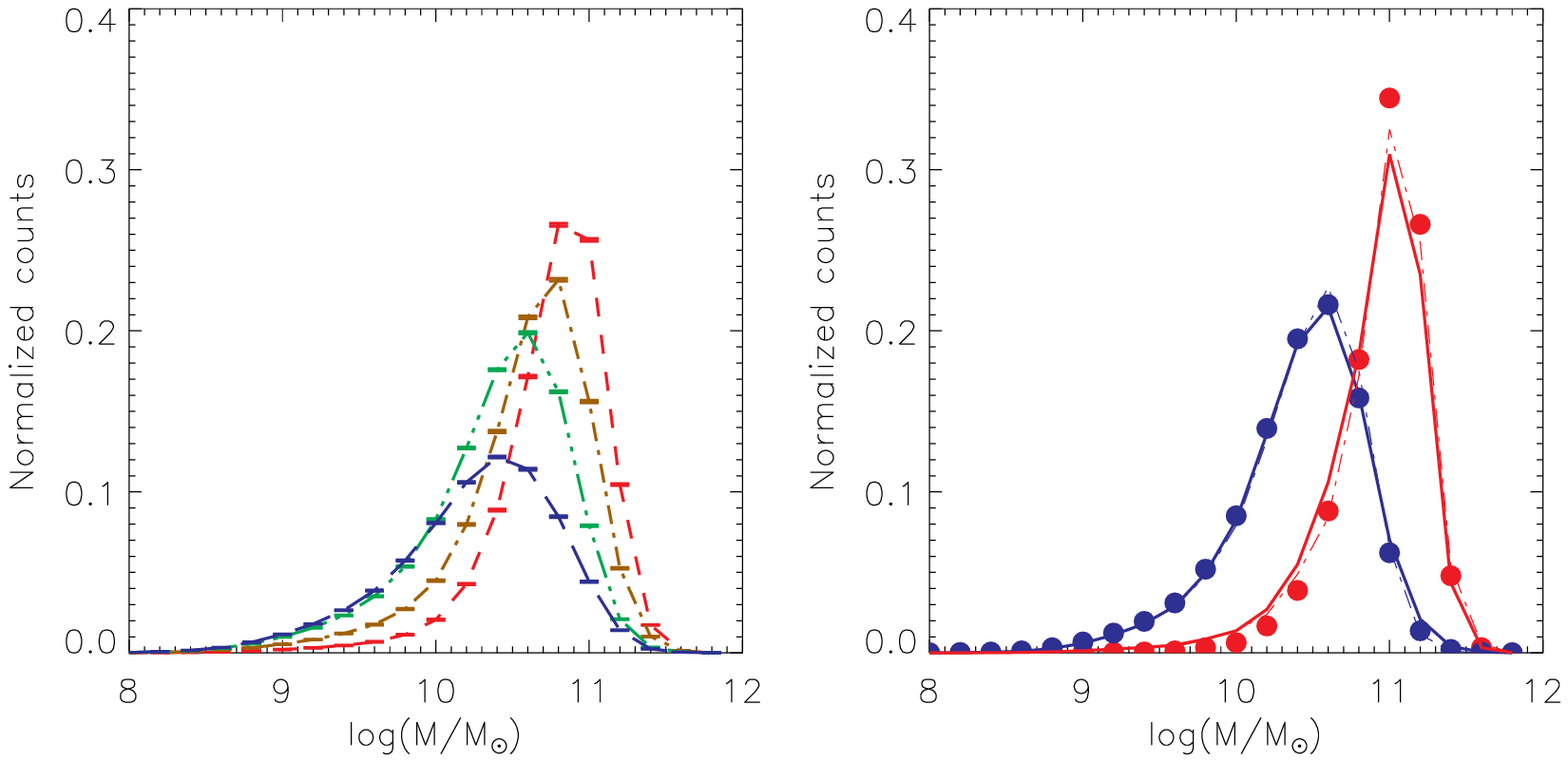}}
 \caption{Observed distribution of masses for different morphological types computed using different estimators described in the text (see text for details). In the left panel the whole sample is shown using the probability weighting. Red short dashed line: ellipticals; yellow dashed dotted line: S0s; green dashed three dotted line: Sabs; blue long dashed line: Scds. In the right panel, we show galaxies flagged as SPIRAL and ELLIPTICAL in the galaxy zoo. Red solid lines are galaxies flagged as ellipticals in Galaxy Zoo (FLAG ELLIPTICAL = 1), red dashed line is the distribution obtained using probability weighting and red dots are galaxies with $p(E)>0.5$. Blue solid lines are galaxies flagged as spirals ( FLAG SPIRAL = 1) in the Galaxy Zoo, blue dashed line is the distribution obtained using probability weighting and blue dots are galaxies with $p(Sab)+P(Scd)>0.5$. }.  
 \label{fig:mass_counts_proba} 
 \end{figure*}

 Another common application is to study the color-stellar mass diagrams for different "robust" morphological types. Again, the probability estimator can be used by computing the 2D histogram of galaxies in the color-mass plane weighted with the probabilities. Figure~\ref{fig:CMR_morpho} shows the probability contours in the color-stellar mass plane for the 4 morphological types. We observe the expected trend: elliptical and S0 galaxies are redder with less scatter, while Sab and Scd are bluer. An interesting feature of Sab galaxies (and for some Scd) is that there seems to be 2 distinct populations: one red population and another one lying in the so-called green valley between the blue cloud and the red sequence . After careful visual inspection of an important fraction of these red galaxies, we can confirm that for most of them they are in fact edge-on spirals probably reddened by dust. A small fraction are, however, \emph{real} passive spirals as shown and carefully studied by \cite{Masters10a, Masters10b}.  Most of them are classified as Sab galaxies with high probability (see figure~\ref{fig:gal_examples}). This result confirms that a pure color selection is not enough to select ellipticals or S0 galaxies since it is highly polluted by edge-on spirals as already shown in previous works (e.g. \citealp{Schawinski07, Lintott08, Bernardi10}). 
 
 These plots are just shown here to validate the morphological classification. A more detailed analysis of the fundamental parameters of galaxies is expected to come in future dedicated papers. 

\begin{figure*} 
 \resizebox{\hsize}{!}{\includegraphics{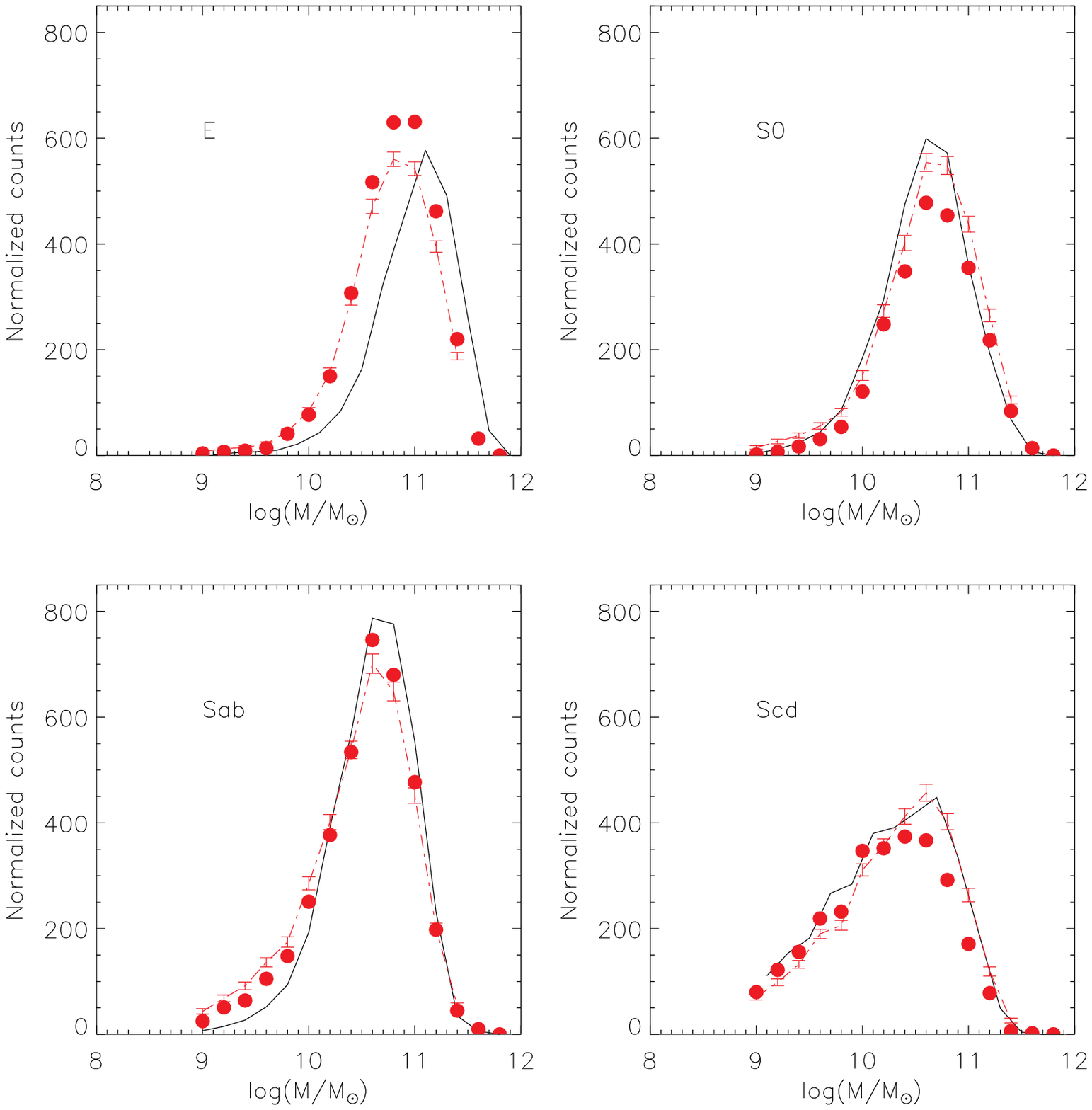}}
 \caption{Observed distribution of masses for different morphological types in the \cite{Nair10} sample using different estimators (see tex for details). Black solid lines: visual classification; red filled circles: probability cuts; red dashed line: probability estimates. Each panel shows a visual morphological class from \cite{Nair10}, selected as described in the text. For the probability cuts, we use $P>0.45$ in this given type. }.  
 \label{fig:mass_counts_nair} 
 \end{figure*}

\begin{figure*} 
 \resizebox{\hsize}{!}{\includegraphics{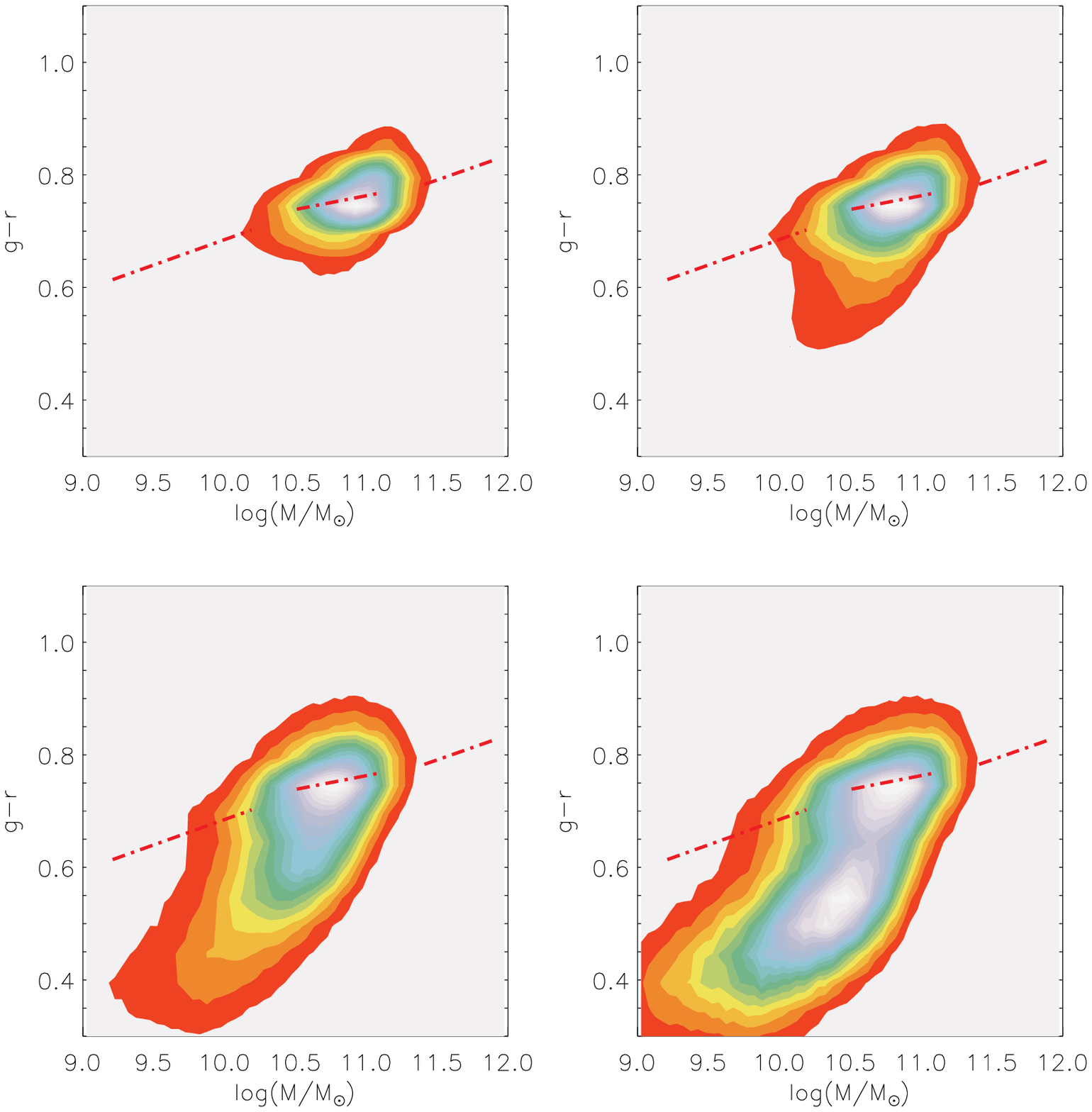}}
 \caption{Color magnitude relation for the 4 morphological types. Contours are computed by probability weighting. For reference, we show in the 4 panels the best fit to the elliptical red sequence from \cite{bernardi10b}. Top left panel: Ellipticals, Top right panel: S0s, bottom left panel: Sab galaxies, bottom right panel: Scd galaxies.}.  
 \label{fig:CMR_morpho} 
 \end{figure*}

\section{Summary and conclusions}

We have presented an automated morphological classification of the SDSS DR7 spectroscopic sample. The algorithm used is based on SVM, and the most interesting and new property is that it associates a probability value to each galaxy instead of a single class. This way, the transition between one class and another is continuous, which should be a better approximation to nature and to visual classifications. As a matter of fact, when the brain decides which morphological class is closer to a given object we are looking at, it probably also implicitly \emph{measures} some parameters and computes distances in this virtual parameter space to decide which one is the closest canonical class to the object it is classifying. In that sense, even if the list of parameters we measure is reduced and much more simplistic than what our brain can do (e.g we are not including spiral arms nor tidal features that certainly play an important role in a visual classification), the spirit of our approach is closer to a classical visual classification than other existing automated methods.  The results obtained are in good agreement with existing visual classifications and are robust even at the faint end of the sample. The main advantage of this approach is that it is fast (a few minutes on a regular laptop) and reproducible. Moreover, we obtain a classification into 4 morphological types instead of the 2 obtained in the Galaxy Zoo. The probability measurements can be used as a weighting factor for computing statistical quantities, such as luminosity or mass functions, or as a selection criterion to be sure that a \emph{cleaned} sample of galaxies is selected. The classification is intended for use in many different applications and is therefore freely available at \url{http://gepicom04.obspm.fr/sdss\_morphology/Morphology\_2010.html}  and soon from the CasJobs database. In subsequent papers, the classification will be used to compare spectroscopic and morphological classifications and investigate possible transitions in color-mass space (Sanchez-Almeida et al. in preparation) and to study the morphological properties of galaxies around BCGs (Bernardi et al. in preparation).

\acknowledgement
The authors are grateful to F. Hammer for reading the manuscript and providing interesting input.

\bibliographystyle{aa}
\bibliography{biblio}

\begin{thebibliography}{37}
\expandafter\ifx\csname natexlab\endcsname\relax\def\natexlab#1{#1}\fi

\bibitem[{Abraham {et~al.}(1996)Abraham, van~den Bergh, Glazebrook, Ellis,
  Santiago, Surma, \& Griffiths}]{abraham96}
Abraham, R., van~den Bergh, S., Glazebrook, K., {et~al.} 1996, ApJ Supplement,
  107, 1

\bibitem[{{Abraham} {et~al.}(1994){Abraham}, {Valdes}, {Yee}, \& {van den
  Bergh}}]{abraham94}
{Abraham}, R.~G., {Valdes}, F., {Yee}, H.~K.~C., \& {van den Bergh}, S. 1994,
  ApJ, 432, 75

\bibitem[{{Abraham} {et~al.}(2003){Abraham}, {van den Bergh}, \&
  {Nair}}]{abraham03}
{Abraham}, R.~G., {van den Bergh}, S., \& {Nair}, P. 2003, \apj, 588, 218

\bibitem[{{Ball} {et~al.}(2004){Ball}, {Loveday}, {Fukugita}, {Nakamura},
  {Okamura}, {Brinkmann}, \& {Brunner}}]{Ball04}
{Ball}, N.~M., {Loveday}, J., {Fukugita}, M., {et~al.} 2004, \mnras, 348, 1038

\bibitem[{{Bell} {et~al.}(2003){Bell}, {McIntosh}, {Katz}, \&
  {Weinberg}}]{Bell03}
{Bell}, E.~F., {McIntosh}, D.~H., {Katz}, N., \& {Weinberg}, M.~D. 2003, \apjs,
  149, 289

\bibitem[{{Bernardi} {et~al.}(2010{\natexlab{a}}){Bernardi}, {Roche},
  {Shankar}, \& {Sheth}}]{bernardi10b}
{Bernardi}, M., {Roche}, N., {Shankar}, F., \& {Sheth}, R.~K.
  2010{\natexlab{a}}, ArXiv e-prints

\bibitem[{{Bernardi} {et~al.}(2010{\natexlab{b}}){Bernardi}, {Shankar}, {Hyde},
  {Mei}, {Marulli}, \& {Sheth}}]{Bernardi10}
{Bernardi}, M., {Shankar}, F., {Hyde}, J.~B., {et~al.} 2010{\natexlab{b}},
  \mnras, 404, 2087

\bibitem[{{Blanton} {et~al.}(2005){Blanton}, {Schlegel}, {Strauss},
  {Brinkmann}, {Finkbeiner}, {Fukugita}, {Gunn}, {Hogg}, {Ivezi{\'c}}, {Knapp},
  {Lupton}, {Munn}, {Schneider}, {Tegmark}, \& {Zehavi}}]{Blanton05}
{Blanton}, M.~R., {Schlegel}, D.~J., {Strauss}, M.~A., {et~al.} 2005, \aj, 129,
  2562

\bibitem[{Chang \& Lin(2001)}]{libSVM}
Chang, C.-C. \& Lin, C.-J. 2001, {LIBSVM}: a library for support vector
  machines, software available at
  \url{http://www.csie.ntu.edu.tw/~cjlin/libsvm}

\bibitem[{{Conselice} {et~al.}(2000){Conselice}, {Bershady}, \&
  {Jangren}}]{Con00}
{Conselice}, C.~J., {Bershady}, M.~A., \& {Jangren}, A. 2000, \apj, 529, 886

\bibitem[{{Conselice} {et~al.}(2008){Conselice}, {Rajgor}, \&
  {Myers}}]{conselice08}
{Conselice}, C.~J., {Rajgor}, S., \& {Myers}, R. 2008, \mnras, 386, 909

\bibitem[{{de Souza} {et~al.}(2004){de Souza}, {Gadotti}, \& {dos
  Anjos}}]{deSouza04}
{de Souza}, R.~E., {Gadotti}, D.~A., \& {dos Anjos}, S. 2004, \apjs, 153, 411

\bibitem[{{Delgado-Serrano} {et~al.}(2010){Delgado-Serrano}, {Hammer}, {Yang},
  {Puech}, {Flores}, \& {Rodrigues}}]{Delgado10}
{Delgado-Serrano}, R., {Hammer}, F., {Yang}, Y.~B., {et~al.} 2010, \aap, 509,
  A78+

\bibitem[{{Emsellem} {et~al.}(2007){Emsellem}, {Cappellari}, {Krajnovi{\'c}},
  {van de Ven}, {Bacon}, {Bureau}, {Davies}, {de Zeeuw}, {Falc{\'o}n-Barroso},
  {Kuntschner}, {McDermid}, {Peletier}, \& {Sarzi}}]{Emsellem07}
{Emsellem}, E., {Cappellari}, M., {Krajnovi{\'c}}, D., {et~al.} 2007, \mnras,
  379, 401

\bibitem[{{Fukugita} {et~al.}(2007){Fukugita}, {Nakamura}, {Okamura}, {Yasuda},
  {Barentine}, {Brinkmann}, {Gunn}, {Harvanek}, {Ichikawa}, {Lupton},
  {Schneider}, {Strauss}, \& {York}}]{fukugita07}
{Fukugita}, M., {Nakamura}, O., {Okamura}, S., {et~al.} 2007, \aj, 134, 579

\bibitem[{Hubble(1926)}]{Hubble26}
Hubble, E.~P. 1926, Astrophys. J., 64, 321

\bibitem[{{Huertas-Company} {et~al.}(2010){Huertas-Company}, {Aguerri},
  {Tresse}, {Bolzonella}, {Koekemoer}, \& {Maier}}]{huertas-company10}
{Huertas-Company}, M., {Aguerri}, J.~A.~L., {Tresse}, L., {et~al.} 2010, \aap,
  515, A3+

\bibitem[{{Huertas-Company} {et~al.}(2009{\natexlab{a}}){Huertas-Company},
  {Foex}, {Soucail}, \& {Pell{\'o}}}]{huertas-company09b}
{Huertas-Company}, M., {Foex}, G., {Soucail}, G., \& {Pell{\'o}}, R.
  2009{\natexlab{a}}, \aap, 505, 83

\bibitem[{{Huertas-Company} {et~al.}(2008){Huertas-Company}, {Rouan}, {Tasca},
  {Soucail}, \& {Le F{\`e}vre}}]{huertas-company08}
{Huertas-Company}, M., {Rouan}, D., {Tasca}, L., {Soucail}, G., \& {Le
  F{\`e}vre}, O. 2008, A\&A, 478, 971

\bibitem[{{Huertas-Company} {et~al.}(2009{\natexlab{b}}){Huertas-Company},
  {Tasca}, {Rouan}, {Pelat}, {Kneib}, {Le F{\`e}vre}, {Capak}, {Kartaltepe},
  {Koekemoer}, {McCracken}, {Salvato}, {Sanders}, \&
  {Willott}}]{huertas-company09}
{Huertas-Company}, M., {Tasca}, L., {Rouan}, D., {et~al.} 2009{\natexlab{b}},
  \aap, 497, 743

\bibitem[{{Kannappan} {et~al.}(2009){Kannappan}, {Guie}, \&
  {Baker}}]{Kannappan09}
{Kannappan}, S.~J., {Guie}, J.~M., \& {Baker}, A.~J. 2009, ArXiv e-prints

\bibitem[{{Kauffmann} {et~al.}(2003){Kauffmann}, {Heckman}, {White}, {Charlot},
  {Tremonti}, {Brinchmann}, {Bruzual}, {Peng}, {Seibert}, {Bernardi},
  {Blanton}, {Brinkmann}, {Castander}, {Cs{\'a}bai}, {Fukugita}, {Ivezic},
  {Munn}, {Nichol}, {Padmanabhan}, {Thakar}, {Weinberg}, \&
  {York}}]{Kauffmann03}
{Kauffmann}, G., {Heckman}, T.~M., {White}, S.~D.~M., {et~al.} 2003, \mnras,
  341, 33

\bibitem[{{Lintott} {et~al.}(2010){Lintott}, {Schawinski}, {Bamford}, {Slosar},
  {Land}, {Thomas}, {Edmondson}, {Masters}, {Nichol}, {Raddick}, {Szalay},
  {Andreescu}, {Murray}, \& {Vandenberg}}]{lintott10}
{Lintott}, C., {Schawinski}, K., {Bamford}, S., {et~al.} 2010, ArXiv e-prints

\bibitem[{{Lintott} {et~al.}(2008){Lintott}, {Schawinski}, {Slosar}, {Land},
  {Bamford}, {Thomas}, {Raddick}, {Nichol}, {Szalay}, {Andreescu}, {Murray}, \&
  {Vandenberg}}]{Lintott08}
{Lintott}, C.~J., {Schawinski}, K., {Slosar}, A., {et~al.} 2008, \mnras, 389,
  1179

\bibitem[{{Masters} {et~al.}(2010{\natexlab{a}}){Masters}, {Mosleh}, {Romer},
  {Nichol}, {Bamford}, {Schawinski}, {Lintott}, {Andreescu}, {Campbell},
  {Crowcroft}, {Doyle}, {Edmondson}, {Murray}, {Raddick}, {Slosar}, {Szalay},
  \& {Vandenberg}}]{Masters10a}
{Masters}, K.~L., {Mosleh}, M., {Romer}, A.~K., {et~al.} 2010{\natexlab{a}},
  \mnras, 405, 783

\bibitem[{{Masters} {et~al.}(2010{\natexlab{b}}){Masters}, {Nichol}, {Bamford},
  {Mosleh}, {Lintott}, {Andreescu}, {Edmondson}, {Keel}, {Murray}, {Raddick},
  {Schawinski}, {Slosar}, {Szalay}, {Thomas}, \& {Vandenberg}}]{Masters10b}
{Masters}, K.~L., {Nichol}, R., {Bamford}, S., {et~al.} 2010{\natexlab{b}},
  \mnras, 404, 792

\bibitem[{{M{\'e}ndez-Abreu} {et~al.}(2008){M{\'e}ndez-Abreu}, {Aguerri},
  {Corsini}, \& {Simonneau}}]{mendez-abreu08}
{M{\'e}ndez-Abreu}, J., {Aguerri}, J.~A.~L., {Corsini}, E.~M., \& {Simonneau},
  E. 2008, \aap, 487, 555

\bibitem[{{Nair} \& {Abraham}(2010)}]{Nair10}
{Nair}, P.~B. \& {Abraham}, R.~G. 2010, \apjs, 186, 427

\bibitem[{{Neichel} {et~al.}(2008){Neichel}, {Hammer}, {Puech}, {Flores},
  {Lehnert}, {Rawat}, {Yang}, {Delgado}, {Amram}, {Balkowski}, {Cesarsky},
  {Dannerbauer}, {Fuentes-Carrera}, {Guiderdoni}, {Kembhavi}, {Liang},
  {Nesvadba}, {{\"O}stlin}, {Pozzetti}, {Ravikumar}, {di Serego Alighieri},
  {Vergani}, {Vernet}, \& {Wozniak}}]{Neichel08}
{Neichel}, B., {Hammer}, F., {Puech}, M., {et~al.} 2008, \aap, 484, 159

\bibitem[{{Peng} {et~al.}(2002){Peng}, {Ho}, {Impey}, \& {Rix}}]{Peng02}
{Peng}, C.~Y., {Ho}, L.~C., {Impey}, C.~D., \& {Rix}, H.-W. 2002, \aj, 124, 266

\bibitem[{{Postman} {et~al.}(2005){Postman}, {Franx}, {Cross}, {Holden},
  {Ford}, {Illingworth}, {Goto}, {Demarco}, {Rosati}, {Blakeslee}, {Tran},
  {Ben{\'{\i}}tez}, {Clampin}, {Hartig}, {Homeier}, {Ardila}, {Bartko},
  {Bouwens}, {Bradley}, {Broadhurst}, {Brown}, {Burrows}, {Cheng}, {Feldman},
  {Golimowski}, {Gronwall}, {Infante}, {Kimble}, {Krist}, {Lesser}, {Martel},
  {Mei}, {Menanteau}, {Meurer}, {Miley}, {Motta}, {Sirianni}, {Sparks}, {Tran},
  {Tsvetanov}, {White}, \& {Zheng}}]{postman05}
{Postman}, M., {Franx}, M., {Cross}, N.~J.~G., {et~al.} 2005, \apj, 623, 721

\bibitem[{{Prieto} {et~al.}(2001){Prieto}, {Aguerri}, {Varela}, \&
  {Mu{\~n}oz-Tu{\~n}{\'o}n}}]{prieto01}
{Prieto}, M., {Aguerri}, J.~A.~L., {Varela}, A.~M., \&
  {Mu{\~n}oz-Tu{\~n}{\'o}n}, C. 2001, \aap, 367, 405

\bibitem[{{S{\'a}nchez Almeida} {et~al.}(2010){S{\'a}nchez Almeida}, {Aguerri},
  {Mu{\~n}oz-Tu{\~n}{\'o}n}, \& {de Vicente}}]{sanchez-almeida10}
{S{\'a}nchez Almeida}, J., {Aguerri}, J.~A.~L., {Mu{\~n}oz-Tu{\~n}{\'o}n}, C.,
  \& {de Vicente}, A. 2010, \apj, 714, 487

\bibitem[{{Schawinski} {et~al.}(2009){Schawinski}, {Lintott}, {Thomas},
  {Sarzi}, {Andreescu}, {Bamford}, {Kaviraj}, {Khochfar}, {Land}, {Murray},
  {Nichol}, {Raddick}, {Slosar}, {Szalay}, {Vandenberg}, \&
  {Yi}}]{Schawinski09}
{Schawinski}, K., {Lintott}, C., {Thomas}, D., {et~al.} 2009, \mnras, 396, 818

\bibitem[{{Schawinski} {et~al.}(2007){Schawinski}, {Thomas}, {Sarzi},
  {Maraston}, {Kaviraj}, {Joo}, {Yi}, \& {Silk}}]{Schawinski07}
{Schawinski}, K., {Thomas}, D., {Sarzi}, M., {et~al.} 2007, \mnras, 382, 1415

\bibitem[{{Simard} {et~al.}(2002){Simard}, {Willmer}, {Vogt}, {Sarajedini},
  {Phillips}, {Weiner}, {Koo}, {Im}, {Illingworth}, \& {Faber}}]{Simard02}
{Simard}, L., {Willmer}, C.~N.~A., {Vogt}, N.~P., {et~al.} 2002, \apjs, 142, 1

\bibitem[{{Trujillo} {et~al.}(2001){Trujillo}, {Aguerri}, {Cepa}, \&
  {Guti{\'e}rrez}}]{trujillo01}
{Trujillo}, I., {Aguerri}, J.~A.~L., {Cepa}, J., \& {Guti{\'e}rrez}, C.~M.
  2001, \mnras, 321, 269

\end{thebibliography}
\appendix
\section{Catalog}
\tiny
\begin{table*}
\caption{First 10 objects in the catalog. Columns are: id, identification number, SpecObjId, id from the SDSS spectroscopic catalog, RA, right ascension, DEC: declination, z, redshift from the SDSS database, p(Early),
probability of being early-type (E or S0), p(E), probability of being elliptical, p(S0), probability of being S0, p(Sab), probability of being Sa or Sb, p(Scd), probability of being Sc or Sd and ask\_class, the spectral class from \cite{sanchez-almeida10} }
\begin{tabular}{cccccc}
\hline\hline\noalign{\smallskip}

  \multicolumn{1}{c}{\emph{id}} &
  \multicolumn{1}{c}{\emph{SpecObjId}} &
  \multicolumn{1}{c}{\emph{RA}} &
  \multicolumn{1}{c}{\emph{DEC}} &
  \multicolumn{1}{c}{\emph{z}} \\
  \multicolumn{1}{c}{\emph{p(Early)}} &
  \multicolumn{1}{c}{\emph{p(E)}} &
  \multicolumn{1}{c}{\emph{p(S0)}} &
  \multicolumn{1}{c}{\emph{p(Sab)}} &
  \multicolumn{1}{c}{\emph{p(Scd)}} &
  \multicolumn{1}{c}{\emph{ask\_class}} \\
\noalign{\smallskip}\hline
  1 & 7509409297491... & 146.7441406 & -0.6522176 & 0.203 \\ \smallskip
   0.941 & 0.790 & 0.150 & 0.032 & 0.026 & 2.0 \\
  2 & 7509409298330... & 146.6285706 & -0.7651463 & 0.064 \\ \smallskip
   0.145 & 0.023 & 0.121 & 0.641 & 0.213 & 0.0 \\
  3 & 7509409301266... & 146.9341278 & -0.670413 & 0.121 \\  \smallskip
   0.969 & 0.861 & 0.108 & 0.016 & 0.013 & 0.0 \\
  4 & 7509409301685... & 146.9638977 & -0.5450143 & 0.056 \\ \smallskip
   0.061 &  0.011 & 0.049 & 0.440 & 0.498 & 10.0 \\
  5 & 7509409302105... & 146.9635162 & -0.7593367 & 0.09 \\ \smallskip
   0.802 & 0.169 & 0.632 & 0.135 & 0.062 & 3.0 \\
  6 & 7509409302524... & 146.9499969 & -0.5922154 & 0.064 \\ \smallskip
  0.120 & 0.020 & 0.100 & 0.762& 0.116 & 10.0 \\
  7 & 7509409303363... & 146.8598328 & -0.8089029 & 0.126 \\ \smallskip
   0.834 & 0.038 & 0.796 & 0.089 & 0.076 & 1.0 \\
  8 & 7509409303783... & 146.5927277 & -0.7602585 & 0.064 \\ \smallskip
   0.188 & 0.026 & 0.161 & 0.618 & 0.193 & 9.0 \\
  9 & 7509409304202... & 146.8576965 & -0.6628734 & 0.084 \\ \smallskip
   0.004 & 0.001 & 0.003 & 0.451 & 0.543 & 9.0 \\
  10 & 7509409304621... & 146.727951 & -0.5568492 & 0.089 \\ \smallskip
   0.939& 0.721 & 0.217 & 0.031 & 0.029 & 0.0 \\
 
\hline
\end{tabular}
\end{table*}

\end{document}